\documentclass[a4paper,11pt]{article}
\usepackage[a4paper,margin=2cm]{geometry}
\usepackage{xcolor}
\usepackage{graphicx}
\usepackage{makecell}
\usepackage{listings}
\usepackage[colorlinks=true]{hyperref}
\usepackage{amsmath}
\usepackage{multirow}
\usepackage{float}
\usepackage{lineno}
\usepackage{subcaption}
\usepackage{pifont}
\usepackage{authblk}
\usepackage{caption}
\usepackage{array}
\begin{document}
\title{Upgrade of the PandaX-4T online data quality monitoring system and perspectives for future multi-tons PandaX upgrades}
\date{}

\def\shKeyLab{INPAC and School of Physics and Astronomy, Shanghai Jiao Tong University, MOE Key Lab for Particle Physics, Astrophysics and Cosmology, Shanghai Key Laboratory for Particle Physics and Cosmology, Shanghai 200240, China}
\def\leeinst{New Cornerstone Science Laboratory, Tsung-Dao Lee Institute, Shanghai Jiao Tong University, Shanghai, 200240, China}
\author[1]{Yubo Zhou}
\author[2,1,3]{Xun Chen\thanks{corresponding author, chenxun@sjtu.edu.cn}}
\affil[ ]{(on behalf of the PandaX-4T Collaboration)}

\affil[1]{\shKeyLab}
\affil[2]{\leeinst}
\affil[3]{Shanghai Jiao Tong University Sichuan Research Institute,\\ Chengdu 610213, China}
\maketitle
\abstract{ PandaX-4T is a xenon-based multi-purpose
  experiment, focusing on particle and astrophysics research.  The data
  quality monitoring system plays a crucial role in the experiment. This system
  enables the prompt detection of potential issues during data
  collection. In order to meet the upgrade requirements of the
  experiment, we have implemented several updates to improve overall
  data throughput and provide users with more comprehensive
  information. As a result, the system is capable of monitoring half
  of the collected data in future operations of the PandaX-4T
  experiment. Furthermore, with updated hardware, the system is also
  well-equipped to meet the requirements of the future multi-ten-tonne
  level PandaX-xT experiment. }

\section{Introduction}

The PandaX-4T experiment~\cite{PandaX:2018wtu} is a multi-tonne liquid
xenon observatory located in the China Jinping Underground Laboratory
~\cite{Li:2014rca}.  It utilizes the dual-phase liquid xenon
technology~\cite{RevModPhys.82.2053} to explore a wide range of
physics phenomena, such as dark matter
detection~\cite{PandaX-4T:2021bab}, coherent elastic scattering of
solar neutrinos~\cite{PandaX:2022aac, Pang:2024bmg}, double beta decay
of xenon isotopes~\cite{PandaX:2022kwg} and other rare
events~\cite{XENON:2019dti}.  After completing the initial data taking
stage (Run1), a detector upgrade was carried out.  The recent
successful completion of the detector upgrade has paved the way for
the scheduling of new data taking stage (Run2).  The future plans
include the development of PandaX-xT, a multi-ten-tonne observatory
that will operate at the same site~\cite{PandaX:2024oxq}.

Scattering events occurring in the active region of the dual-phase
xenon detector may produce prompt scintillation light signal ($S1$) in liquid
xenon and electroluminescence signal ($S2$) in gas xenon. These signals can be
captured by two arrays of photomultiplier tubes (PMTs) on the top and
bottom of the detector, read out separately with 169 and 199 channels. 
The captured signals are digitized using flash
analog-to-digital converters (FADCs) and subsequently stored for
further analysis.  Due to the combination of high digitization rate of 
250~MHz (500~MHz post-upgrade~\cite{He:2021sbc}) and a
large number of readout channels, a substantial amount of data is
generated. It is estimated that PandaX-4T is collecting approximately 1
PB of data per year~\cite{Zhou:2023vmz}, while PandaX-xT is projected
to produce over 1.8 PB of data per year~\cite{PandaX:2024oxq}. These data
are stored in a custom format based on the Bamboo Shoot3
library~\cite{PandaX:bamboo-shoot3}. During the data taking process,
an Apache Kafka message queue~\cite{Kreps2011KafkaA,Sax2018,Apache:Kafka} is employed to transmit
metadata information to various functional modules. These modules
encompass the online data processing, data transfering and data quality
monitoring system.

The data quality monitoring system plays a crucial role in the data
taking process by offering reference plots and numerical
values from the collected data. These resources serve as valuable
references for both the on-site operators and the data analyzers. 
In conjunction with the upgrade of the
current PandaX-4T experiment, the doubled sampling rate results in a
higher data flow. This poses challenges for the existing data quality
monitoring system to extract reference resources in a timely
manner. To address this, we have performed a comprehensive upgrade of
the system, making it more suitable to meet the requirements of both
the upcoming operation of PandaX-4T and the future PandaX-xT
experiment.

This paper provides a comprehensive description of the
architecture and upgrade of the data quality monitoring system. In
Sec.~\ref{sec:framework} we present an introduction to the data
architecture of the system, together with its
sub-modules. Sec.~\ref{sec:upgrade_plotg} focuses on the upgrade of
the data quality information extraction, aimed at improving data
throughput. In Sec.~\ref{sec:web_upgrade}, we introduce an
enhancement of the web-based user interface, which involves a precise
presentation of data. Finally, a summary is
provided in Sec.~\ref{sec:summary}.

\section{The data quality monitoring system}
\label{sec:framework}

In this section, we introduce the architecture, components, and
functionalities of the data quality monitoring system.

\subsection{Architecture of the system}
\label{sec:architecture}

\begin{figure}[hbt]
  \centering
  \includegraphics[width=0.9\textwidth]{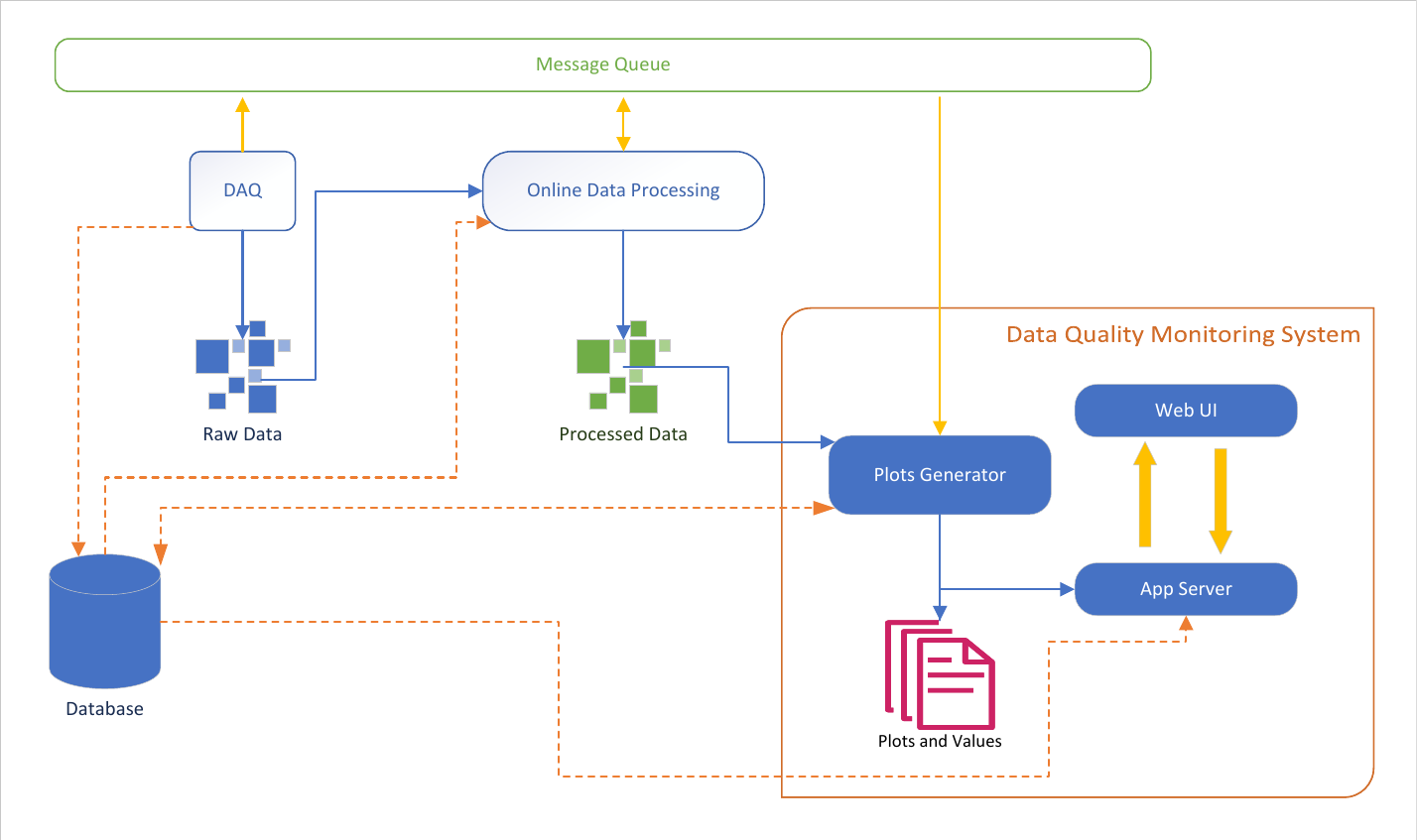}
  \caption{ Architecture of the data quality monitoring system as well
    as the online data processing system. The blue lines indicate the
    direction of the data flow. The dashed orange lines in the diagram
    represent the information exchange with the database, while the
    solid yellow lines represent the information exchange with the
    message queue. The online data processing system processes
    specific raw data produced by the data acquisition (DAQ) program. The plots generator
    extracts reference quality information from the processed data like radon event rate and electron lifetime, 
    which is then saved as plots and data files or stored in the
    database. Users can access this information through the WebUI,
    which interacts with the app server to retrieve the required files
    or data from the database.}
  \label{fig:dq_framework}
\end{figure}

The data quality monitoring system is essential for swiftly
identifying issues during data collection. It must extract physical
information from processed data obtained through the online data
processing system once the data is ready. Furthermore, the metadata of
runs is stored in the database. Thus, the data quality monitoring
system is tightly integrated with the online data processing system
and the database, as illustrated in Fig.~\ref{fig:dq_framework}. The
system comprises three independent sub-modules: a) the plots
generator, responsible for extracting information from processed
files; b) the Web UI, which serves as the user interface; c) the app
server, reading extracted information from the database or file system
and transmitting it to the user. The interactions among these
sub-modules are distinctly represented in Fig.~\ref{fig:dq_framework}.

The different systems within the architecture are interconnected using
the Apache Kafka message queue. This message queue allows for the
creation and sharing of different ``topics'', which serve as
categories for organizing and sharing messages among the various
systems involved~\cite{Kreps2011KafkaA,Sax2018,Apache:Kafka}.  In the
DAQ system, raw data from a single run are stored as a series of
files, each assigned a unique file number for identification.  After
completing the writing process for a data file, a message containing
the run and file number is sent to the ``daq'' topic. The online data
processing system consumes this message and determines whether or not
to process the raw data based on predefined rules. The data processing
steps employed by the system follow the standard procedures introduced
in Ref.~\cite{Zhou:2023vmz}. Once the processing of a given file is
completed, the same message is sent to the ``dq''(data quality) topic
to notify the plots generator about extracting the data quality
information.  In the previous version of the plots generator, about 50
image files were generated for different variables that are valuable
for quality monitoring, and some information is stored directly to the
database.

\begin{figure}[hbt]
  \centering
  \includegraphics[width= 0.8 \textwidth]{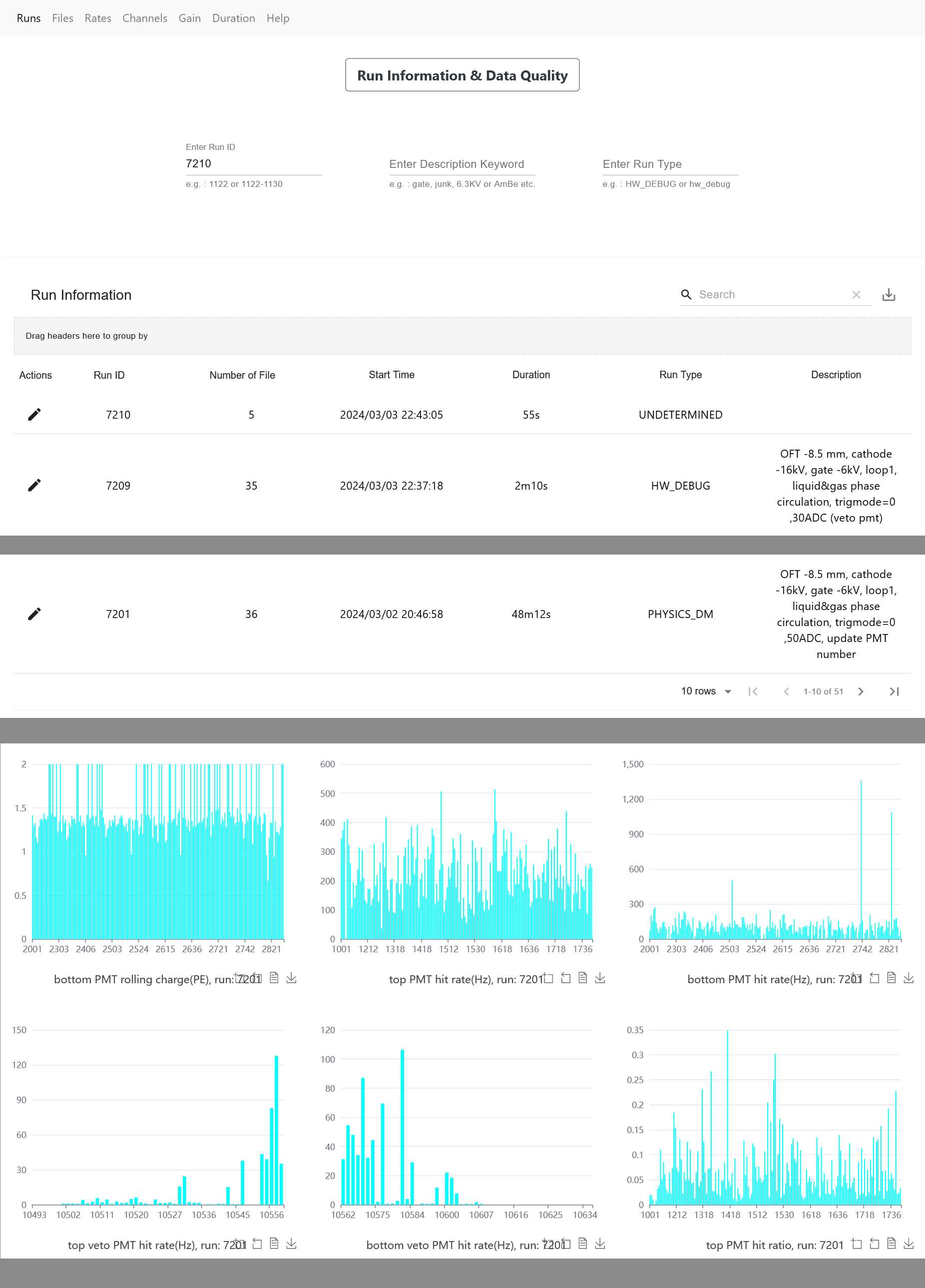}  
  \caption{Screenshot of the interface displaying run-related data
    quality information. (Note: Image is modified to fit the
    page. Grayed-out sections have been omitted.)}
  \label{fig:run_interface}
\end{figure}

To access the plots and related information, users can utilize their
browsers to visit the WebUI, which is served by a Nginx web server
~\cite{Nginx}. The WebUI, developed using the React
library~\cite{Facebook:React}, initially displays the run list. When a
user selects a specific run from the interface, the corresponding data
quality plots and values are fetched from the back-end app server and
displayed below the run list. The example screenshot is shown in
Fig.~\ref{fig:run_interface}.

The back-end app server is built using the lightweight Echo
framework~\cite{LabStack:Echo}. It offers representational state
transfer (REST) application programming interfaces (APIs) that allow
users to access the related data quality information.

\subsection{Plots generator}

For each selected raw data file, the online data processing system
generates three output files, containing hit, signal, and event level
information, respectively.  A hit corresponds to a pulse in the raw
waveform of a PMT, containing information about detected photons. A
signal is a grouping of hits from multiple PMTs with similar
timing. An event consists of signals occurring within a specific time
window, potentially originating from the same physical
interaction~\cite{Zhou:2023vmz}.  Once the processing of a data file
is completed, the plots generator is notified through the message
queue. The multi-threaded plots generator operates with one thread
dedicated to receiving messages and placing them into an internal data
structure, which is a mapping between run numbers and the list of
corresponding file numbers.  Upon the arrival of a message for a new
run, a dedicated thread is assigned to generate data quality plots for
that particular run. The thread analyzes the output files in the order
specified in the corresponding file number lists. Once all the related
files have been analyzed, the file number is removed from the list,
and the files are subsequently removed from storage. The thread is
recycled for a new run when the file list remains empty for a minimum
of one hour.

Presently, more than 50 distributions about data quality variables are
derived from the processed data. From the files containing hit level
information, variables such as hit rate, noise-to-signal pulse ratio,
stability of channel baseline, and hit areas are extracted. The status
of PMTs is monitored using the collected charges of each PMT based on 
the signal data. Additionally, the rate of small $S1$ signals and
sparking signals is monitored to aid identifying potential detector
malfunctions. Various signal properties are collected for distribution
analysis. At the event level, the lightened PMTs of events with single $S1$
or single $S2$ signals are monitored, along with other types of
special events. For the ``good'' events that consist of one major $S1$
followed by one major $S2$, the top-bottom asymmetry of the $S1$
signal and the time separation between S1 and S2 signal serve as
references for evaluating the detector status. Several variables,
such as the rates of radon events and low energy electron recoil
events, and the electron lifetime are calculated.

In the previous version of the plots generator, the distributions were
stored in memory using \texttt{TH1D} or \texttt{TH2D} objects from the
ROOT~\cite{TEAM:ROOT} framework. These objects were updated as the
input files were analyzed. The plots were regenerated in PNG format
with updated information after processing every 10 raw files. The
calculated variables were simultaneously uploaded to the database.

\subsection{WebUI}

The WebUI presents information in six categories across different
pages, including runs, files, channels, rates, PMT gains, and
durations. Most of the data quality information is specific to each
run. By default, the interface exhibits a list of the 50 latest
runs. Above this list, searching boxes are provided to enable users to
search for specific runs based on the run number, run number range,
keywords in the description, or the type. Each entry in the list
corresponds to a run and can be edited by the shifter to update the
run type and description. Upon clicking on a run entry, the data
quality plots and numbers corresponding to that particular run will be
displayed below the run list.

The page for displaying file information primarily serves the purpose
of manually tagging abnormal or bad files. On the other hand, the page
dedicated to displaying channel information is utilized for tagging
bad channels.  The rates page offers a comprehensive view of the evolution 
of key variables over time, including the electron lifetime, hit
rates, signal rates, and rates of specific events. These
variables serve as references for assessing the long-term
stability of the detector.  The gains page allows users to examine the
gain evolution of each PMT and explore the gain pattern for specific
runs. The duration page offers a comprehensive view of the accumulated 
run duration of different types of data. It displays the duration of the 
user-input run list, excluding any files that have been tagged as bad. Within
these pages, the data are visualized in the browser using the EChartsJS
package~\cite{Apache:ECharts}.

\subsection{App server}

The app server has the responsibility of reading or updating the
database, searching for related files in the storage, and sending the
results back to the user through the reverse proxy provided by the
Nginx HTTP server. The Go-Pg library is utilized for accessing the
PostgreSQL database.  As the system is designed solely for internal
use, there are no authentication and logging functionalities
implemented in the app server.  All the functionalities of the app
server are exposed as REST APIs. The APIs are summarized in Table
~\ref{tab:apis}.

\begin{table}[hbt]
  \centering
  \begin{tabular}{l|c|l}
    \hline
    API & method & description \\\hline
    \texttt{/runs} & GET & get list of runs \\
    \texttt{/run/image/:id} & GET & get images of a run with given id \\
    \texttt{/run/run\_rates/:id} & GET & get rates of a run with given id \\
    \texttt{/run/trigger\_rates/:id} & GET & get trigger rates of a run with given id \\
    \texttt{/run/board\_rates/:id} & GET & get board rates of a run with given id \\
    \texttt{/run/gain/:id} & GET & get PMT gains of a run with given id \\
    \texttt{/runs/duration/type} & GET & get duration of run of given types \\
    \texttt{/runs/duration/sum} & GET & get duration of run within a given list \\
    \texttt{/run/:id} & PUT & update run description and type\\
    \texttt{/run/channels} & PUT & set bad channels \\
    \texttt{/files} & GET & get list of files \\
    \texttt{/files/update} & PUT & update file quality information \\\hline
  \end{tabular}
  \caption{The APIs provided by the app server.}
  \label{tab:apis}
\end{table}

\section{Upgrade of the plots generator}
\label{sec:upgrade_plotg}

The updated plots generator offers a significant increase in
throughput compared to the initial version. The updated plots
generator also includes support for the prompt acquisition of
self-corrected PMT gain by utilizing online generated hit data.

\subsection{Performance optimization}

The first version of plots generator was designed to handle input data
sequentially. It has been effective in processing data with a low data
rate, such as background data, by processing one file out of every two
files. However, during a high bandwidth calibration run,
later files experience longer waiting times for processing. To avoid
the blocking, the fraction of processed files was reduced to 1/8 in Run0 
and Run1 calibration process, but it may result in the generated plots 
being less representative of the collected data.

Another concern with the current plots generator is the generation of
images using the ``\texttt{Print()}'' function of \texttt{TH1D} or
\texttt{TH2D} classes from ROOT, which resulting in performance
downgrade. On the server equipped with two AMD EPYC 7452 CPUs and 256
GB of memory, the operation to generate all the required data quality
images took approximately 3.76 seconds on average.  Additionally, as
the number of data runs increases, the accumulation of small image
files becomes a burden on the file system and difficult to manage.

In the upcoming upgrade of PandaX-4T and future
PandaX-xT experiments, the data rates are expected to be significantly
higher. To meet the new demand based on the previously mentioned 
drawbacks, two major updates have been implemented.

At first, we made an update by replacing the internal data structure
to store data quality information from \texttt{TH1D} and \texttt{TH2D}
to vectors in the C++ standard library. To improve efficiency, we have
decided not to generate image files. Instead, we
now dump the data quality data of a run into a single text file in
JSON format.  For each run, the text file that contains all the data
quality information has a size of approximately 7 MB.

Another significant optimization involves utilizing all available
threads in the pre-allocated thread pool to concurrently process
incoming files. This allows files with different file numbers to be
processed in parallel, improving overall efficiency. Before updating
the vectors used for storing data quality information, the working
thread must acquire the corresponding lock to prevent data racing.

The overall improvement in processing speed is studied with
the PandaX-4T data from various types of runs. To check the
performance, a thread pool with only one thread is used, as 
in the previous implementation of the program. The results are presented in
Tab.~\ref{tab:performance_one}. It can be observed that the throughput
has improved by over 50\% with only one working thread.

\begin{table}[hbt]
  \centering
  \begin{tabular}{c|cc|cc|c}
    \hline
    \multirow{2}{*} {Run type} &  \multicolumn{2}{c|}{processing time (minutes)} & \multicolumn{2}{c|}{throughput (MB/s)} & \multicolumn{1}{c} {improvement} \\
    &  previous & updated & previous & updated & in throughput\\\hline
    background &  194.15 & 119.37 & 19.34 & 31.45 & 62.6\% \\
    Rn & 201.43 & 123.93 & 18.64 & 30.30 & 64.3\% \\
    $^{83\rm{m}}$Kr & 191.35 & 118.73 & 19.62 & 31.62 & 61.2\% \\
    DD & 213.13 & 130.42 & 17.62 & 28.79 & 63.4\%\\
    $^{241}$AmBe & 203.28 & 132.25 & 18.47 & 28.39 & 53.7\% \\
    $^{232}$Th & 209.1 & 132.00 & 17.96 & 28.44 & 58.4\% \\
    $^{60}$Co & 215.23 & 135.17 & 17.44 & 27.78 & 59.2\% \\\hline
  \end{tabular}
  \caption{Comparison of processing time and throughput for two
    versions of plots generators on various types of data. For each
    run, the initial 200 files are processed. Each file has a size of
    approximately 1.1~GB. The updated version utilizes only one thread
    for data processing.}
  \label{tab:performance_one}
\end{table}

\begin{figure}[hbt]
  \centering
  \includegraphics[width=0.6\textwidth]{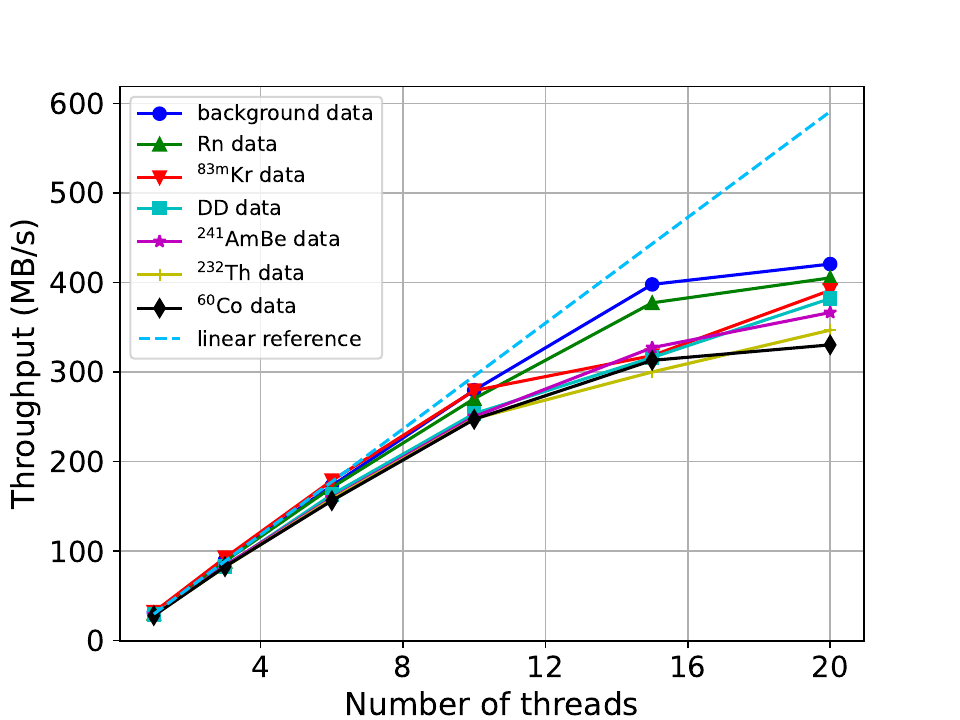}  
  \caption{Throughput (MB/s) of the plots generator with allocated
    number of threads for different types of data. The dashed blue  
    line indicates the linear reference.}
  \label{fig:throughput_multi_threads}
\end{figure}

By increasing the number of threads in the pre-allocated thread pool,
the throughput of the updated plots generator exhibits nearly linear
improvement initially. However, after surpassing 10 threads, the
improvement becomes less pronounced, as depicted in
Fig.~\ref{fig:throughput_multi_threads}. This behavior can be attributed
to the constraints imposed by concurrent reading from the storage and
the need to maintain data integrity through locking mechanisms in a
multi-threaded environment.

Given the constraint of limited available CPU cores on the server, a
careful consideration was made to strike a balance between data
processing and plot generation. Ultimately, a decision was made to use 
4 threads in the pool. This configuration allows for
efficient data monitoring while simultaneously accommodating online
data processing. With this setup, the plots generator achieves a final
throughput of 100-120 MB/s. Considering that the data based on the
Bamboo Shoot3 library is already compressed, resulting in
approximately 45\% space savings, this throughput can be translated to
around 180-220 MB/s for the raw binary data from the DAQ. Based on
previous experience with the operation of PandaX-4T, the raw data rate
for background runs is approximately 30 MB/s, while for Rn calibration
runs it is around 160 MB/s. With the new DAQ system's doubled sampling
rate, the raw data rate will double as well, reaching 320 MB/s. Selecting 
4 threads enables the online analysis of half of the generated data, providing 
valuable insights into the data quality.  To meet the demands of
future experiments, upgrading the server with additional CPU cores
would allow for more threads, and consequently a higher throughput, 
addressing the need for increased data processing capacity.

\subsection{Support for the extraction of the PMT gain}

The PandaX-4T experiment utilizes weekly calibration with
light-emitting diodes (LEDs) to calculate PMT gains. These LEDs,
driven by a standard driver circuitry with adjusted voltage, can
generate fast and faint light pulses with appropriate
occupancies~\cite{Kapustinsky:1985hnc, Li:2015qhq}. This allows for
the clean separation of single photoelectrons (SPEs) which are used to
fit the PMT gains, also known as LED gains. However, this method may
encounter challenges such as uneven illumination and insufficient SPE
samples~\cite{PandaX:2023ggs}. To address these issues, a recent
analysis of PandaX-4T~\cite{Luo:2024rud} adopts SPEs from physics data
to correct the PMT gains, known as self-corrected gains, based on the
calculated LED gains.

The self-corrected gains of the PMTs were extracted from every 1 in 100
files during a massive data reprocessing.  This was accomplished by
identifying the single photon hits within the selected data, and
fitting their charge distribution. The charge values used for the
fitting were obtained with the LED gains. In an ideal case, it was
expected that the fitted peak with Gaussian function would be at 1 
photoelectron (PE). The fitted value is used to correct the PMT gain. 
The corrected values are stored in the database and utilized as 
self-corrected gains for the hit charges during the subsequent analyses. 

The inclusion of self-corrected gains extraction in the offline data
processing chain resulted in a 17\% increase in processing time for
the complete processing of one raw file. This increase is due to the
need to obtain information from a newly generated file with hit
information, which is skipped in the default offline data
processing. In contrast, the hit file is generated by default in
online processing, as it is required by the plots generator to
calculate hit-related variables. The update adding the functionality
of extracting self-corrected gain values to the plots generator did
not result in a noticeable increase in processing time.  The benefits
of this update extend beyond time-saving measures. By leveraging the
plots generator, lower statistical errors in the determination of
self-corrected gains can be achieved compared to the previous methods,
as more data files will be processed.  Additionally, this update aids
in the identification of potential issues with PMTs during data
collection.

\section{Upgrade of the web application}
\label{sec:web_upgrade}
Incorporating JSON-based quality information requires an update to the
WebUI to enable the direct visualization of this data in the
browser. The JSON file organizes different data categories as an array
of mappings, each of which corresponds to a specific name and its
associated data. The WebUI utilizes the \texttt{EChartJS} package
to generate plots for each element in the array. The current WebUI
predominantly employs three types of plots: bar charts, heatmaps, and
scatter plots.

\begin{figure}[hbt]
  \centering
  \begin{subfigure}[b]{0.3291\textwidth}
    \includegraphics[width=\textwidth]{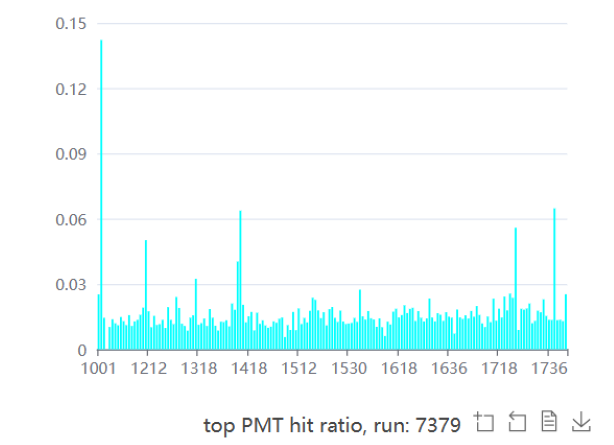}
    \caption{}
    \label{fig:hit_level_hit_ratio}
  \end{subfigure}
  \begin{subfigure}[b]{0.3254\textwidth}
    \includegraphics[width=\textwidth]{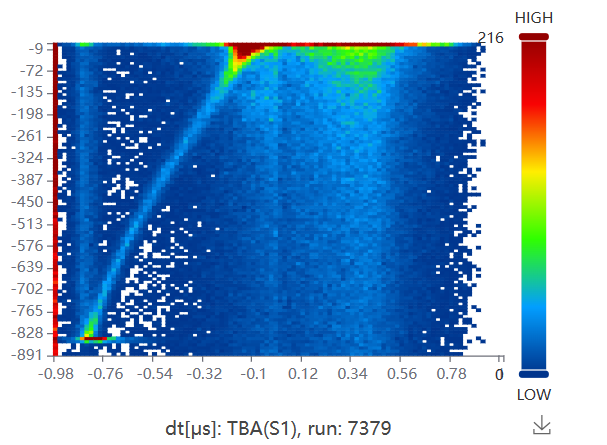}
    \caption{}
    \label{fig:physical_level_dt_s1_tba}
  \end{subfigure}
  \begin{subfigure}[b]{0.2455\textwidth}
    \includegraphics[width=\textwidth]{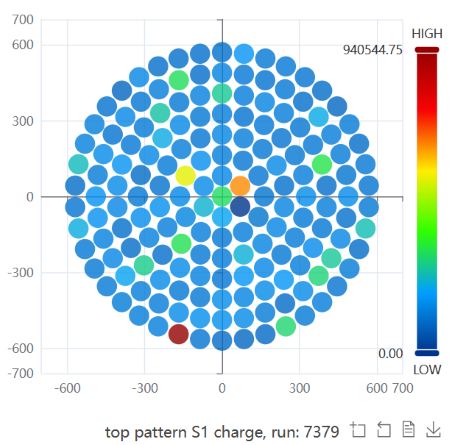}
    \caption{}
    \label{fig:physical_level_top_s1_charge}
  \end{subfigure}
  \caption{Examples of bar chart(\subref{fig:hit_level_hit_ratio}),
    heatmap(\subref{fig:physical_level_dt_s1_tba}) and scatter 
    plot(\subref{fig:physical_level_top_s1_charge}).}
  \label{fig:plots_types}
\end{figure}

Bar charts are utilized as replacements for \texttt{TH1D} histograms,
particularly for channel-based data quality information. Heatmaps
serve as substitutes for \texttt{TH2D} histograms, allowing for the
visualization of two-dimensional distributions. Scatter plots are
primarily employed to display PMT-related information. Examples of
these different plot types are depicted in
Figure~\ref{fig:plots_types}. These plots can be easily exported as
SVG files. Additionally, a new function has been added to the
scatter plots, allowing users to view the detailed information of each data point directly.

Plotting multiple images simultaneously in the browser can lead to
significant performance issues. To address this, we have used the 
lazy plotting technique, with which the plotting of images is
deferred until they are within the browser window, ensuring efficient
resource utilization.

To ensure the continuous monitoring of data quality information during
an ongoing data collection run, a new auto-refreshing function has
been implemented. This function automatically requests new data every
30 seconds and updates the related plots accordingly. This feature
proves to be extremely valuable for shifters, as it allows them to stay up-to-date with the
latest data quality information.

Minor updates have been made to the back-end app server as well, including
the addition of two REST APIs. These APIs enable users to obtain
the JSON information of a run and retrieve the self-corrected gain
information.

In addition to the aforementioned modifications, the configuration of
the Nginx web server has been updated to enable gzip compression for
text data, which automatically compresses the JSON file
during network transfer. As a result, a 7 MB JSON file is compressed to
approximately 200 KB, significantly reducing network bandwidth
consumption.

\section{Summary}
\label{sec:summary}
The data quality monitoring system plays a crucial role in ensuring 
prompt detection of potential data-related issues during the PandaX-4T 
data collection. In this paper, we have presented the framework of 
the data quality monitoring system in the PandaX-4T experiment,  
and most importantly the upgrades made to its critical components. 

Regarding the plots generator, we have made several updates. Firstly,
we have revamped the internal data structure by replacing the
histogram classes from ROOT with standard containers. Secondly, we
have implemented multi-threading processing of incoming files,
incorporating a locking mechanism to maintain data integrity during
concurrent updating of the run level information. Additionally, we
have transitioned from using image files to using JSON files for
extracting data quality information. These updates have significantly
improved the overall data throughput, allowing us to monitor half of
the collected data online in the future operations of the PandaX-4T
experiment.

As for the WebUI, our update focuses on client-side plotting based on
JSON data obtained from the app server. This update reduces the
requirement of network bandwidth and provides users with more
extensive information compared to the previous implementation.

The updated data quality monitoring system successfully fulfills all
necessary requirements for the upcoming operation of the PandaX-4T
experiment. Furthermore, it demonstrates the potential to effectively
monitor data with higher bandwidth in the future PandaX-xT experiment,
especially when running on a more powerful server.

\section*{Acknowledgments}
\label{sec:ack}
This project is supported by the grants from National Natural Science
Foundation of China (No. 12175139).

\section*{Conflict of Interest Statement}
On behalf of all authors, the corresponding author states that there is no conflict of interest.

\bibliographystyle{hunsrt}
\bibliography{references/refs}

\end{document}